\documentclass[a4paper,11pt]{article}
\usepackage{pos}
\usepackage{slashed}

\title{Semileptonic kaon decays and the precise determination of $V_{us}$}

\author*[a,b]{Chien-Yeah Seng}

\affiliation[a]{Facility for Rare Isotope Beams, Michigan State University, \\
East Lansing, MI 48824, USA}

\affiliation[b]{Department of Physics, University of Washington,\\
Seattle, WA 98195-1560, USA}

\emailAdd{seng@frib.msu.edu}

\renewcommand{\logo}{\relax}

\abstract{I will give a brief overview of the Standard Model theory inputs needed for the precise determination of the Cabibbo-Kobayashi-Maskawa matrix element $V_{us}$ from semileptonic kaon decays, focusing on the long-distance electromagnetic corrections. I will then describe our recent effort to pin down this correction at sub-permille level, which further sharpens the so-called ``Cabibbo angle anomaly'', an interesting observation which may point towards new physics.}

\FullConference{The 11th International Workshop on Chiral Dynamics (CD2024)\\
 26-30 August 2024\\
Ruhr University Bochum, Germany\\}

\begin{document}
\maketitle

\section{Introduction}

The elements of the Cabibbo-Kobayashi-Maskawa (CKM) matrix~\cite{Cabibbo:1963yz,Kobayashi:1973fv}
\begin{equation}
	V_{\text{CKM}}=\left(\begin{array}{ccc}
		V_{ud} & V_{us} & V_{ub}\\
		V_{cd} & V_{cs} & V_{cb}\\
		V_{td} & V_{ts} & V_{tb}
	\end{array}\right)~,
\end{equation}
that describe the mixing of quark flavor eigenstates, are among the basic parameters of the Standard Model (SM), and the precise determination of their values provide a useful avenue to test the validity of the SM and to search for physics beyond the SM (BSM). The unitarity of $V_\text{CKM}$ is a definite prediction of SM that can be tested against experiment. For instance, it implies the following relation between the first-row matrix elements:
\begin{equation}
	|V_{ud}|^2 + |V_{us}|^2 + |V_{ub}|^2=1~.\label{eq:unitarity}
\end{equation}
Currently, the relation above is tested at the level of $0.01\%$. Given that $|V_{ub}|^2\sim 10^{-5}$~\cite{ParticleDataGroup:2024cfk} is an order of magnitude smaller than even the current uncertainties of both $|V_{ud}|^2$ and $|V_{us}|^2$, dropping it  leads to a simpler ``Cabibbo unitarity'' $V_{ud}^2+V_{us}^2=1$ or, in other words, $V_{ud}=\cos\theta_C$, $V_{us}=\sin\theta_C$ where $\theta_C$ is the Cabibbo angle. Assuming new physics occurs at a heavy scale $\Lambda_\text{BSM}$ and affects low-energy ($E\ll v_\text{H}$ with $v_\text{H}\approx 246$~GeV the Higgs vacuum expectation value) charged weak interactions through dimension-six operators, a simple dimensional analysis shows that a $0.01\%$ precision probes new physics at the scale
\begin{equation}
	\left(\frac{v_\text{H}}{\Lambda_\text{BSM}}\right)^2\sim 0.01\% \implies \Lambda_\text{BSM}\sim 20~\text{TeV}~,
\end{equation}
which is competitive to high-energy collider experiments. 

The matrix elements $V_{ud}$ and $V_{us}$ can be extracted from charged weak decays of hadrons and nuclei. In particular, the best determination of $V_{ud}$ (at face value) comes from superallowed $0^+\rightarrow 0^+$ decay of $T=1$ nuclei~\cite{Hardy:2020qwl}, whereas $V_{us}$ is best determined from semileptonic decays of kaon ($K_{\ell 3}$). Additionally, the ratio $V_{us}/V_{ud}$ is best determined from the ratio of leptonic decays of kaons ($K_{\mu 2}$) and pions ($\pi_{\mu 2}$). Given the central value $V_{ud}^2\sim 0.95$ and $V_{us}^2\sim 0.05$, to test Eq.\eqref{eq:unitarity} at $0.01\%$ level requires the relative precision of the extraction of $V_{ud}$ and $V_{us}$ to reach $0.01\%$ and $0.1\%$, respectively. This imposes challenges not only on the experimental side, but also on the theory side because one needs to understand all the SM effects in these decay processes to the desired level of precision, and for that one needs to overcome the large theory uncertainties from the non-perturbative Quantum Chromodynamics (QCD) that binds quarks and gluons into hadrons.

In this talk I discuss the SM corrections to semileptonic decays of kaons, with particular emphasis on long-distance radiative corrections (RC), including previous state-of-the-art computation with chiral perturbation theory (ChPT) and our recent improvement based on a novel, hybrid formalism. I will conclude by quoting the newly-determined $V_{us}$ and discussing the current status of the first-row CKM unitarity.

\section{SM theory inputs to semileptonic kaon decays}

There are six independent channels of $K_{\ell 3}$ decays, all with experimentally measured lifetime and branching ratio: $K^L\rightarrow \pi^- e^+\nu_e$ ($K_{e3}^L$), $K^L\rightarrow \pi^-\mu^+\nu_\mu$ ($K_{\mu 3}^L$)~\cite{KLOE:2005vdt,KLOE:2005lau,Vosburgh:1972zqy,KTeV:2004hpx}, $K^S\rightarrow \pi^- e^+\nu_e$ ($K_{e3}^S$), $K^S\rightarrow \pi^- \mu^+\nu_e$ ($K_{\mu 3}^S$)~\cite{KTeV:2010sng,KLOE:2010yit,NA48:2002iol,Bertanza:1996dt,Schwingenheuer:1995uf,Gibbons:1993zj,Batley:2007zzb,KLOE:2006vvm,KLOE:2002lao}, $K^+\rightarrow \pi^0 e^+\nu_e$ ($K_{e3}^+$), and $K^+\rightarrow \pi^0 \mu^+\nu_e$ ($K_{\mu 3}^+$)~\cite{KLOE:2007wlh,Koptev:1995je,Ott:1971rs,Lobkowicz:1969mx,Fitch:1965zz,KLOE:2007jte,Chiang:1972rp}. Here I define $K_{\ell 3}$ to be inclusive, namely it may include radiations of arbitrary number of real photons. From these decays, one extracts $V_{us}^2$ through the following master formula of the partial decay rate $\Gamma_{K_{\ell 3}}$~\cite{ParticleDataGroup:2024cfk}:
\begin{equation}
	\Gamma_{K_{\ell 3}}=\frac{G_F^2V_{us}^2M_K^5C_K^2}{192\pi^3}S_\text{EW}|f_+^{K^0\pi^-}(0)|^2 I_{K\ell}^{(0)}\left(1+\delta_{\text{SU(2)}}^{K\pi}+\delta_\text{EM}^{K\ell}\right)~,\label{eq:master}
\end{equation}
where $G_F=1.1663787(6)\times 10^{-5}$~GeV$^{-2}$ is Fermi's constant obtained from the muon lifetime~\cite{MuLan:2012sih}, and $C_K$ is an isospin factor that equals 1 ($1/\sqrt{2}$) for $K^0$ ($K^+$) decay. There are a number of SM inputs appearing at the right hand side of Eq.\eqref{eq:master}, which we briefly discuss as follows.

\subsection{$S_\text{EW}$}

The quantity $S_\text{EW}$ encodes the process-independent electroweak RC not included in definition of $G_F$, with the following schematic expression~\cite{Cirigliano:2011ny}:
\begin{equation}
	S_\text{EW}=1+\frac{2\alpha}{\pi}\left(1-\frac{\alpha_s}{4\pi}\right)\ln\frac{M_Z}{M_\rho}+\mathcal{O}\left(\frac{\alpha\alpha_s}{\pi^2}\right)~.
\end{equation}
In practice, one often adopts the numerical value $S_\text{EW}=1.0232(3)$~\cite{Marciano:1993sh,Erler:2002mv}.

\subsection{$f_+^{K^0\pi^-}(0)$}

The matrix element of the charged weak current $J_W^{\mu\dagger}$ between $K$ and $\pi$ can be parameterized as
\begin{equation}
	\langle \pi(p_\pi)|J_W^{\mu\dagger}(0)|K(p_K)\rangle =V_{us}C_K\left[(p_K+p_\pi)^\mu f^{K\pi}_+(t)+(p_K-p_\pi)^\mu f^{K\pi}_-(t)\right]~,
\end{equation}
which defines the form factor $f_{\pm}^{K\pi}(t)$, with $t\equiv (p_K-p_\pi)^2$ the squared momentum transfer between $K$ and $\pi$ (it is also customary to define $f_0^{K\pi}(t)\equiv f_+^{K\pi}(t)+t f_-^{K\pi}(t)/(M_K^2-M_\pi^2)$). In particular, the $t\rightarrow 0$ limit (which is, of course, unphysical) of the $K^0\rightarrow \pi^-$ matrix element gives rise to the quantity $f_+^{K^0\pi^-}(0)$, which is close to but not exactly 1 due to the approximate SU(3) flavor symmetry. It is a fully non-perturbative quantity that must be computed by lattice QCD to per mille level in order to satisfy the required precision goal. The Flavor Lattice Averaging Group (FLAG)~\cite{FlavourLatticeAveragingGroupFLAG:2024oxs} gives the following averaged value~\cite{Carrasco:2016kpy,Bazavov:2018kjg,Bazavov:2012cd,Boyle:2015hfa}:
\begin{eqnarray}
	f_+^{K^0\pi^-}(0) & = & \left\{ \begin{array}{ccc}
		0.9698(17) &  & N_{f}=2+1+1\\
		0.9677(27) &  & N_{f}=2+1
	\end{array}\right.~.\label{eq:latticef}
\end{eqnarray}
On the other hand, the contribution of $f_-^{K\pi}$ to the squared amplitude is suppressed by $m_\ell^2/M_K^2$, so the precision requirement of this form factor is lower. 

\subsection{$I_{K\ell}^{(0)}$} 

The tree-level decay phase space integral, which encodes the $t$-dependence of the decay form factors, is expressed as the following quantity~\cite{Antonelli:2010yf,Cirigliano:2011ny}\footnote{Notice that these references have typos in their expression of $I_{K\ell}^{(0)}$.}:
\begin{equation}
	I_{K\ell}^{(0)}=\int_{m_\ell^2}^{(M_K-M_\pi)^2}\frac{dt}{M_K^8}\bar{\lambda}^{3/2}\left(1+\frac{m_\ell^2}{2t}\right)\left(1-\frac{m_\ell^2}{t}\right)^2\left[\bar{f}_+^2(t)+\frac{3m_\ell^2\Delta_{K\pi}^2}{(2t+m_\ell^2)\bar{\lambda}}\bar{f}_0^2(t)\right]~,
\end{equation}
with $\bar\lambda\equiv [t-(M_K+M_\pi)^2][t-(M_K-M_\pi)^2]$ and $\Delta_{K\pi}\equiv M_K^2-M_\pi^2$; we have also introduced the rescaled form factors, $\bar{f}(t)\equiv f^{K\pi}(t)/f_+^{K\pi}(0)$. 

\begin{table}
	\begin{centering}
		\begin{tabular}{cc}
			\hline 
			Mode & $I_{K\ell}^{(0)}$\tabularnewline
			\hline 
			$K_{e3}^{0}$ & 0.15470(15)\tabularnewline
			\hline 
			$K_{e3}^{+}$ & 0.15915(15)\tabularnewline
			\hline 
			$K_{\mu3}^{0}$ & 0.10247(15)\tabularnewline
			\hline 
			$K_{\mu3}^{+}$ & 0.10553(16)\tabularnewline
			\hline 
		\end{tabular}
		\par\end{centering}
	\caption{The tree-level $K_{\ell3}$ phase-space integral with the most recent dispersive parameterization
		of form factors~\cite{,PSCKM21}.\label{tab:PSfactor}}
	
\end{table}

One way to obtain the $t$-dependence of the decay form factors is to fit the experimental Dalitz plot to specific parameterizations of the form factors, for example the Taylor expansion~\cite{Antonelli:2010yf}, the $z$-parameterization~\cite{Hill:2006bq}, the pole parameterization~\cite{Lichard:1997ya} and the dispersive parameterization~\cite{Bernard:2006gy,Bernard:2009zm,Abouzaid:2009ry}. Currently the dispersive parameterization, which result is given in Table~\ref{tab:PSfactor}, has claimed the smallest uncertainty. Another possibility is to study the $t$-dependence through lattice QCD.

\subsection{$\delta_{\text{SU(2)}}^{K\pi}$}
 
Next, we have the isospin breaking correction factor $\delta_\text{SU(2)}^{K\pi}$ defined as:
\begin{equation}
	\delta_\text{SU(2)}^{K\pi}\equiv\left(\frac{f_+^{K\pi}(0)}{f_+^{K^0\pi^-}(0)}\right)^2-1
\end{equation}
which resides only in the $K^+_{\ell 3}$ channel. Upon neglecting small electromagnetic contributions, it depends on two combinations of quark mass parameters:
$m_s/\hat{m}$ and $(m_s^2-\hat{m}^2)/(m_d^2-m_u^2)$, where $\hat{m}=(m_u+m_d)/2$. These parameters can be obtained either from lattice QCD~\cite{RBC:2014ntl,Durr:2010vn,Durr:2010aw,MILC:2009ltw,Fodor:2016bgu,Bazavov:2017lyh,EuropeanTwistedMass:2014osg,FermilabLattice:2014tsy,Giusti:2017dmp} or phenomenology ($\eta\rightarrow 3\pi$)~\cite{Colangelo:2018jxw}, however the two return somewhat different results for the isospin breaking correction~\cite{Seng:2021nar}:
\begin{equation}
	\left(\delta_\text{SU(2)}^{K^+\pi^0}\right)_\text{lattice}=0.0457(20)~,~\left(\delta_\text{SU(2)}^{K^+\pi^0}\right)_\text{pheno}=0.0522(34)~,
\end{equation}
which reason is yet to be understood.

\subsection{$\delta_\text{EM}^{K\ell}$}

\begin{figure}
	\centering
	\includegraphics[width=0.8\columnwidth]{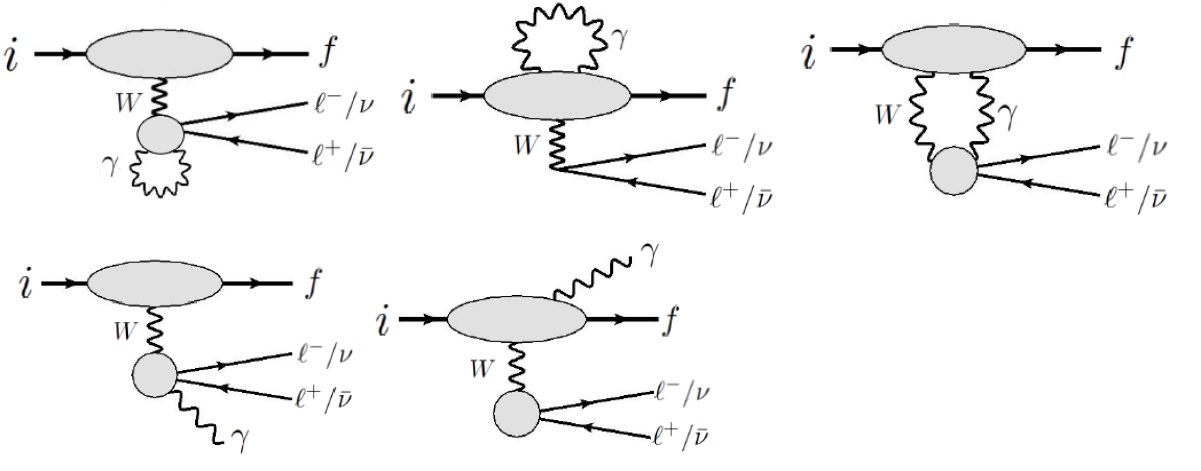}
	\caption{$\mathcal{O}(G_F\alpha)$ diagrams with virtual and real photons.\label{fig:EMdiagrams}} 
\end{figure} 

As the last important SM input, $\delta_\text{EM}^{K\ell}$ denotes the ``long-distance'' electromagnetic (EM) corrections to the decay rate. Schematically, Feynman diagrams in Fig.\ref{fig:EMdiagrams} represent the $\mathcal{O}(G_F\alpha)$ corrections to the decay amplitude involving virtual and real photons, and ``long distance'' means the contribution from these diagrams at the energy scale where the W-boson propagator effectively shrinks to a point (Fermi's interaction). In more rigorous terms, $\delta_\text{EM}^{K\ell}$ simply represents all the remaining electroweak RC that is neither reabsorbed into $G_F$ nor included in $S_\text{EW}$. 

While the bremsstrahlung (real photon emission) contribution is more straightforwardly calculable, one-loop diagrams are more complicated because virtual photons with the loop momentum $q$ probe the hadron physics at all scales, including $q\sim 1$~GeV where QCD becomes fully non-perturbative. The rest of this talk will be dedicated to the discussion of this correction.

\section{ChPT evaluation of $\delta_\text{EM}^{K\ell}$}

ChPT is a low-energy effective field theory (EFT) of the strong interaction, constructed from the spontaneously-broken chiral symmetry of QCD. Its Lagrangian consists of infinitely many terms, arranged in an increasing power of a small chiral expansion parameter (will be discussed below), such that to any given level of precision, only a finite number of terms in the Lagrangian need to be included. Terms in the chiral Lagrangian are accompanied by parameters known as low-energy constants (LECs); they are non-perturbative quantities which values are not constrained by the chiral symmetry, and thus must be deduced either from experiments or computed with first-principles methods such as lattice QCD. 

To describe $K_{\ell 3}$ decays and their RC, the degrees of freedom (DOFs) of the chiral Lagrangian are chosen to be the pseudoscalar meson octet, leptons and photons. We can split the full chiral Lagrangian into:
\begin{equation}
	\mathcal{L}=\mathcal{L}_\text{lepton}+\mathcal{L}_\gamma+\mathcal{L}_\text{ChPT}~,
\end{equation}
where
\begin{eqnarray}
\mathcal{L}_\text{lepton}&=&\sum_\ell [\bar{\ell}(i\slashed{\partial}+e\slashed{A}-m_\ell)\ell+\bar{\nu}_{\ell L}i\slashed{\partial}v_{\ell L}]\nonumber\\
\mathcal{L}_\gamma&=&-\frac{1}{4}F_{\mu\nu}F^{\mu\nu}-\frac{1}{2\xi}(\partial\cdot A)^2+\frac{1}{2}M_{\gamma}^2A_\mu A^\mu
\end{eqnarray}
are the pure leptonic and photonic pieces, respectively (with a small fictitious photon mass $M_\gamma$ to regularize the infrared (IR)-divergence). To allow a simultaneous expansion of $\mathcal{L}_\text{ChPT}$ in both the small momentum and EM coupling, we define a chiral expansion parameter $\epsilon$ that scales as:
\begin{equation}
	\epsilon\sim p/\Lambda_\chi\sim e~,
\end{equation}
where $p$ is a typical external momentum of the process, $\Lambda_\chi\sim 1$~GeV is the so-called ``chiral symmetry breaking scale'' that signifies the onset of non-perturbative effects, and $e$ is the Quantum Electrodynamics (QED) coupling constant. With this we can write:
\begin{equation}
	\mathcal{L}_\text{ChPT}=\mathcal{L}^{(2)}+\mathcal{L}^{(4)}+\dots
\end{equation}
The leading-order (LO) chiral Lagrangian $\mathcal{L}^{(2)}$ consists of two terms that scale as $\mathcal{O}(p^2)$ and $\mathcal{O}(e^2)$ respectively~\cite{Gasser:1984gg}:
\begin{equation}
	\mathcal{L}^{p^2}=\frac{F_0^2}{4}\langle D_\mu U(D^\mu U)^\dagger+\chi U^\dagger+U\chi^\dagger\rangle~,~\mathcal{L}^{e^2}=ZF_0^4\langle q_L U^\dagger q_R U\rangle~.
\end{equation}
Meanwhile, to compute the long-distance EM correction up to $\mathcal{O}(G_F\alpha)$ requires terms in $\mathcal{L}^{(4)}$ that scale as $\mathcal{O}(p^4,e^2p^2)$:
\begin{equation}
	\mathcal{L}^{(4)}=\mathcal{L}^{p^4}+\mathcal{L}^{e^2p^2}_{\{K\}}+\mathcal{L}^{e^2p^2}_{\{X\}}+\dots
\end{equation}
where $\mathcal{L}^{e^2p^2}_{\{K\}}$ contains dynamical photon fields (with LECs labeled as $\{K_i\}$)~\cite{Urech:1994hd}, while $\mathcal{L}^{e^2p^2}_{\{X\}}$ contains dynamical photon \textit{and} lepton fields (with LECs labeled as $\{X_i\}$)~\cite{Knecht:1999ag}. 
 
\begin{figure}
	\centering
	\includegraphics[width=0.8\columnwidth]{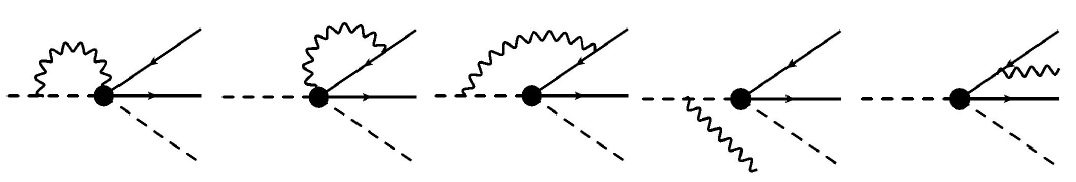}
	\caption{$\mathcal{O}(e^2p^2)$ diagrams with virtual and real photons in ChPT (self-energy diagrams not displayed).\label{fig:ChPTdiagrams}} 
\end{figure}  

In ChPT power counting, the tree-level charged weak matrix element scale as $\mathcal{O}(p^2)$, and thus the leading EM corrections scale as $\mathcal{O}(e^2p^2)$, which come from the loop + bremsstrahlung diagrams in Fig.\ref{fig:ChPTdiagrams} together with the tree-level matrix element of $\mathcal{L}^{e^2p^2}$; the latter is required to cancel the ultraviolet (UV)-divergences from the loops. The values of the renormalized LECs $\{K_i^r\}$ and $\{X_i^r\}$ were estimated with vector dominance models~\cite{Ananthanarayan:2004qk,DescotesGenon:2005pw}; in fact, they encode physics not only at the hadronic scale but also at the weak scale. In particular, the combination~\cite{DescotesGenon:2005pw}
\begin{equation}
	X_6^\text{phys}\equiv X_6^r-4K_{12}^r
\end{equation}
can be split into the long distance (LD) and short distance (SD) part, and the latter simply gives rise to the universal electroweak factor $S_\text{EW}$ in the previous section~\cite{Cirigliano:2002ng}:
\begin{equation}
	S_\text{EW}=1-e^2(X_6^\text{phys})_\text{SD}~.
\end{equation}
 
\begin{table}
 	\begin{centering}
 		\begin{tabular}{cc}
 			\hline 
 			Mode & $\delta_\text{EM}^{K\ell}(\%)$\tabularnewline
 			\hline 
 			$K_{e3}^0$ & $0.99(19)_{e^2p^4}(11)_\text{LEC}$\tabularnewline
 			\hline 
 			$K_{e3}^+$ & $0.10(19)_{e^2p^4}(16)_\text{LEC}$\tabularnewline
 			\hline 
 			$K_{\mu 3}^0$ & $1.40(19)_{e^2p^4}(11)_\text{LEC}$\tabularnewline
 			\hline 
 			$K_{\mu 3}^+$ & $0.02(19)_{e^2p^4}(16)_\text{LEC}$\tabularnewline
 			\hline 
 		\end{tabular}
 		\par\end{centering}
 	\caption{Results of long-distance EM corrections in ChPT~\cite{Cirigliano:2008wn}.\label{tab:deltaEMChPT}}
 	
\end{table}
 
The results of the ChPT evaluation of $\delta_\text{EM}^{K\ell}$ are summarized in Table~\ref{tab:deltaEMChPT}~\cite{Cirigliano:2008wn}. There are two main sources of uncertainties:
\begin{enumerate}
	\item Neglected terms at the order $\mathcal{O}(e^2p^4)$, which uncertainties are obtained by multiplying the central value of the $\mathcal{O}(e^2p^2)$ result by $\epsilon^2\approx M_K^2/\Lambda_\chi^2$.
	\item The poorly-known LECs, which uncertainties are simply taken as 100\% of their size.  
\end{enumerate} 
As a combined result, the ChPT evaluation of $\delta_\text{EM}^{K\ell}$ has an absolute uncertainty at the level $10^{-3}$, which marginally satisfies the precision requirement for the first-row CKM unitarity test. To improve upon that requires: (1) New theory framework to effectively resum the most important $\mathcal{O}(e^2p^n)$ contributions missed in the fixed-order ChPT calculation, and (2) Appropriate lattice QCD inputs to reduce uncertainties from the LECs. 

A possible future path is to compute the full decay amplitude, including the loop and bremsstrahlung diagrams, with lattice QCD; this was successfully performed in $K_{\mu 2}$ decays which results in a reduction of uncertainties from electromagnetic and isospin-breaking corrections~\cite{Giusti:2017dwk}. However, a similar calculation for $K_{\ell 3}$ is expected to be much more challenging, with a projected timeline of $\sim$ 10 years to reach a $10^{-3}$ precision that we need~\cite{BoyleSnowmass}. 

\section{Sirlin's representation} 

\begin{figure}
	\centering
	\includegraphics[width=0.8\columnwidth]{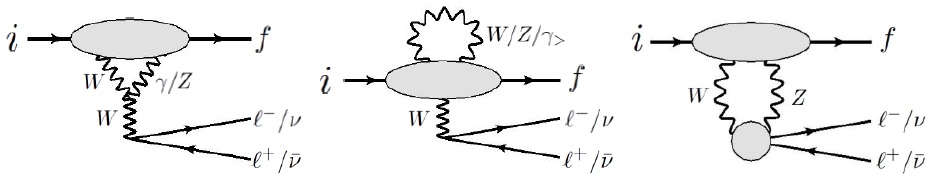}
	\caption{$\mathcal{O}(G_F\alpha)$ ``weak'' RC diagrams.\label{fig:WeakRC}} 
\end{figure} 

To overcome the limitations in ordinary ChPT calculations, one may turn to a classical representation by Sirlin~\cite{Sirlin:1977sv} (I will call it ``Sirlin's representation'') which was first constructed to deal with the $\mathcal{O}(G_F\alpha)$ electroweak RC in superallowed beta decays, and was later re-introduced to study the electroweak RC in general semileptonic decays~\cite{Seng:2019lxf,Seng:2021syx}. It is based on the current algebra relation, namely the equal-time commutation relation between electroweak currents:
\begin{equation}
	[J_W^{0\dagger}(\vec{x},t),J_\text{em}^\mu(\vec{y},t)]=J_W^{\mu\dagger}(\vec{x},t)\delta^{(3)}(\vec{x}-\vec{y})~,~\label{eq:CA}
\end{equation}
which is an exact mathematical identity. I refer interested readers to more comprehensive reviews for details~\cite{Sirlin:1977sv,Seng:2021syx}, but simply state the most important results here. First, within the formalism, the prescription used to differentiate the ``weak'' and ``EM'' RC is the following splitting of the photon propagator (in Feynman gauge):
\begin{equation}
	\frac{1}{q^{\prime 2}}=\frac{1}{q^{\prime 2}-M_W^2}+\frac{M_W^2}{M_W^2-q^{\prime 2}}\frac{1}{q^{\prime 2}}~.
\end{equation}
One sees that, the first term at the right hand side resembles a heavy propagator with mass $M_W$, while the second term is a massless propagator accompanied by a Pauli-Villars regulator. I will denote the corresponding propagator line of these two terms in a Feynman diagram as $\gamma_>$ and $\gamma_<$, respectively. With this, one can identify a subset of the full one-loop Feynman diagrams, namely those in Fig.\ref{fig:WeakRC}, as the ``weak'' RC diagrams, as at $\mathcal{O}(G_F\alpha)$ their corresponding loop integrals are only sensitive to loop momenta of the order $q^\prime\sim M_W$; this is region the QCD coupling is small and the loop integrals are perturbatively calculable. Their effects are either reabsorbed into the definition of $G_F$, or contribute to a finite correction to the tree-level amplitude. Subsequently, we can write the full Lagrangian as:
\begin{equation}
	\mathcal{L}=\mathcal{L}_\text{QCD}+\mathcal{L}_{\text{QED},\gamma_<}+\mathcal{L}_{4f}'~,
\end{equation}
where the second term at the right hand side is the usual QED Lagrangian, but with the photon propagator accompanied by $M_W^2/(M_W^2-q^{\prime 2})$; the third term is a ``modified'' Fermi interaction, which reads:
\begin{equation}
	\mathcal{L}_{4f}'=-\left\{1-\frac{\alpha}{2\pi}\left[\ln\frac{M_W^2}{M_Z^2}+\mathcal{O}(\alpha_s)\right]\right\}\frac{G_F}{\sqrt{2}}J_W^\mu\bar{\ell}\gamma_\mu(1-\gamma_5)\nu+h.c.~.\label{eq:Leff}
\end{equation} 

Denoting the tree-level weak current matrix element as $F^\mu\equiv \langle \phi_f(p_f)|J_W^{\mu\dagger}(0)|\phi_i(p_i)\rangle$ (we concentrate on semileptonic decays with $\ell^{+}$-emission), we can now write the decay amplitude including the $\mathcal{O}(G_F\alpha)$ loop corrections as:
\begin{eqnarray}
	\mathcal{M}&=&\sqrt{Z_\ell}\left[1-\frac{\alpha}{2\pi}\left(\ln\frac{M_W^2}{M_Z^2}+\mathcal{O}(\alpha_s)\right)\right]\mathcal{M}_0-\frac{G_F}{\sqrt{2}}\delta F^\mu L_\mu +\delta \mathcal{M}_{\gamma W}~,
\end{eqnarray} 
where $L_\mu=\bar{u}_\nu\gamma_\mu(1-\gamma_5)v_\ell$ is the leptonic matrix element of the weak current, $\mathcal{M}_0=-(G_F/\sqrt{2})F^\mu L_\mu$ is the tree-level amplitude, and $Z_\ell$ is the known EM-induced $\ell$-wavefunction renormalization factor. There are only two non-trivial quantities in the expression above: (1) $\delta F^\mu$ which denotes the EMRC to the weak matrix element $F^\mu$ (the second diagram in Fig.\ref{fig:EMdiagrams}, with $\gamma\rightarrow\gamma_<$), and (2) $\delta \mathcal{M}_{\gamma W}$ which denotes the $\gamma W$-box diagram (the third diagram in Fig.\ref{fig:EMdiagrams}). 

Using the on-mass-shell perturbation theory~\cite{Brown:1970dd} and the current algebra relation~\eqref{eq:CA}, one can split $\delta F^\mu$ into two pieces:
\begin{equation}
	\delta F^\mu=\delta F^\mu_2+\delta F^\mu_3~,
\end{equation}
which I will call ``two-point function'' and ``three-point function'' respectively (and their correction to the decay amplitude $\delta \mathcal{M}_{2(3)}$). The full dependence on hadron physics in both $\delta F_2^\mu$ and $\delta\mathcal{M}_{\gamma W}$ resides in the so-called ``generalized Compton tensor'':
\begin{equation}
	T^{\mu\nu}(q';p_f,p_i)\equiv \int d^4xe^{iq'\cdot x}\langle \phi_f(p_f)|T\{J_\text{em}^\mu(x)J_W^{\nu\dagger}(0)\}|\phi_i(p_i)\rangle~,
\end{equation}
which involves a time-ordered product of the EM and weak current. One may also define another quantity $\Gamma^\mu(q';p_f,p_i)$ by simply replacing $J_W^{\nu\dagger}$ in $T^{\mu\nu}$ by $\partial\cdot J_W^\dagger$. The full expression of $T^{\mu\nu}$ requires exact knowledge of non-perturbative QCD; however, one can isolate from it an analytic piece known as the ``convection term''~\cite{Meister:1963zz}:
\begin{equation}
	T^{\mu\nu}_\text{conv}(q';p_f,p_i)=\frac{iZ_f(2p_f+q')^\mu F^\nu(p_f,p_i)}{(p_f+q')^2-M_f^2}+\frac{iZ_i(2p_i-q')^\mu F^\nu(p_f,p_i)}{(p_i-q')^2-M_i^2}~,
\end{equation}
which represents the simplest structure satisfying the exact EM Ward identity, and is hence giving rise to the full IR-divergent structure.

The three-point function, on the other hand, depends on a three-current product:
\begin{equation}
	\delta F_3^\mu\sim \int d^4x d^4ye^{iq'\cdot x}\langle \phi_f(p_f)|T\{J_W^{\mu\dagger}(x)J^\nu_\text{em}(y)J^\text{em}_\nu(0)\}|\phi_i(p_i)\rangle~,
\end{equation} 
and vanishes identically if the two conditions below are simultaneously satisfied: (1) The weak current is conserved, i.e. $\partial\cdot J_W=0$, and (2) The decay is forward, i.e. $p_i=p_f$. We can also split the $\gamma W$-box diagram into two pieces:
\begin{equation}
	\delta\mathcal{M}_{\gamma W}=\delta\mathcal{M}_{\gamma W}^a+\delta\mathcal{M}_{\gamma W}^b~,
\end{equation}
where $\delta\mathcal{M}_{\gamma W}^b$ contains an $\epsilon$-tensor that comes from $L_\mu$ while $\delta\mathcal{M}_{\gamma W}^a$ does not. There is a partial cancellation between $\delta F_2^\mu$ and $\delta\mathcal{M}_{\gamma W}^a$ which gets rids of some dependence on non-perturbative QCD. 

With all the above, the virtually-corrected decay amplitude in Sirlin's representation take the following form:
\begin{eqnarray}
	\mathcal{M}&=&\mathcal{M}_0\left\{1+\frac{\alpha}{2\pi}\left[\ln\frac{M_Z^2}{m_\ell^2}-\frac{1}{4}\ln\frac{M_W^2}{m_\ell^2}+\frac{1}{2}\ln\frac{m_\ell^2}{M_\gamma^2}-\frac{3}{8}+\frac{1}{2}\tilde{a}_g\right]+\frac{1}{2}\delta_\text{HO}^\text{QED}\right\}\nonumber\\
		&&+\left(\delta\mathcal{M}_2+\delta\mathcal{M}_{\gamma W}^a\right)_\text{int}-\frac{G_F}{\sqrt{2}}\delta F_3^\lambda L_\lambda+\delta\mathcal{M}_{\gamma W}^b~,\label{eq:Sirlinmain}
\end{eqnarray}
In the first line, $\tilde{a}_g$ represents the $\mathcal{O}(\alpha_s)$ perturbative Quantum Chromodynamics (pQCD) corrections from all one-loop integrals except $\delta\mathcal{M}_{\gamma W}^b$; $\delta_\text{HO}^\text{QED}$ represents the resummation of the leading higher-order QED corrections. Both quantities are process-independent and known numerically.
Meanwhile,  $\left(\delta\mathcal{M}_2+\delta\mathcal{M}_{\gamma W}^a\right)_\text{int}$ represents a ``residual integral'' after removing the exactly calculable piece in the sum of $\delta\mathcal{M}_2$ and $\delta\mathcal{M}_{\gamma W}^a$; this integral is insensitive to the physics at the UV scale.

\section{Application to $K_{\ell 3}$}

We used Eq.\eqref{eq:Sirlinmain} as a starting point to perform a re-analysis of the EM corrections to $K_{\ell 3}$ decay rate~\cite{Seng:2019lxf,Seng:2020jtz,Seng:2021boy,Seng:2021wcf,Seng:2022tjh,Seng:2022wcw}. First, it is always possible to recast the loop corrections to the $K_{\ell 3}$ decay in terms of corrections to the tree-level form factors:
\begin{equation}
	f_{\pm}^{K\pi}(t)\rightarrow f_{\pm}^{K\pi}(t)+\delta f_\pm^{K\pi}(y,z)~,
\end{equation}
where $y\equiv 2p_i\cdot p_\ell/M_K^2$, $z\equiv 2p_i\cdot p_f/M_K^2$. Similar to tree level, the contribution of $\delta f_-^{K\pi}$ to the squared amplitude is suppressed by $m_\ell^2/M_K^2$, so its precision requirement is lower than $\delta f_+^{K\pi}$. 

In Eq.\eqref{eq:Sirlinmain}, there are only three quantities that require non-perturbative inputs: $\left(\delta\mathcal{M}_2+\delta\mathcal{M}_{\gamma W}^a\right)_\text{int}$, $\delta\mathcal{M}_{\gamma W}^b$ and $\delta F_3^\lambda$. We compute their contribution to $\delta f_\pm^{K\pi}$ as follows.

\subsection{$\delta f_+^{K\pi}$}

\begin{figure}
	\centering
	\includegraphics[width=0.5\columnwidth]{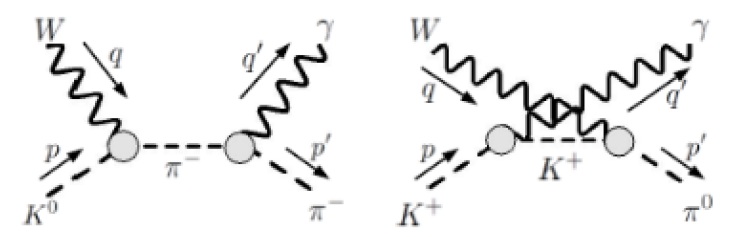}
	\caption{``Pole'' contributions to the generalized Compton tensors.\label{fig:pole}} 
\end{figure} 

We split the full $\delta f_+^{K\pi}$ into three pieces; the first comes from the ``residual integral'' and the part of $\delta\mathcal{M}_{\gamma W}^b$ that involves the vector (V) component of $J_W^{\mu\dagger}$. It reduces to the extraction of $\delta f_+^{K\pi}$ from the following integral:
\begin{eqnarray}
	I_\mathfrak{A}^\lambda&\equiv&-e^2\int\frac{d^4q'}{(2\pi)^4}\frac{1}{[(p_\ell-q')^2-m_\ell^2][q^{\prime 2}-M_\gamma^2]}\left\{\frac{2p_\ell\cdot q'q^{\prime \lambda}}{q^{\prime 2}-M_\gamma^2}T^\mu_{\:\:\mu}+2p_{\ell\mu}T^{\mu\lambda}\right.\nonumber\\
	&&\left.-(p_i-p_f)_\mu T^{\lambda\mu}+i\Gamma^\lambda-i\epsilon^{\mu\nu\alpha\lambda}q_\alpha'(T_{\mu\nu})_V\right\}~,
\end{eqnarray}
which is saturated by the ``pole'' diagram contribution to the generalized Compton tensors, see Fig.\ref{fig:pole}. The inputs here are therefore the $K/\pi$ electromagnetic and charged weak form factors, which are readily obtainable from experiment~\cite{Amendolia:1986ui,Amendolia:1986wj,NA482:2018rgv}. Notice that the use of full form factors effectively resum the most important $\mathcal{O}(e^2p^n)$ corrections in ChPT.

Next, we have the contribution from the part of $\delta\mathcal{M}_{\gamma W}^b$ that involves the axial (A) component of $J_W^{\mu\dagger}$. The corresponding integral reads:
\begin{equation}
	I_\mathfrak{B}^\lambda\equiv ie^2\int\frac{d^4q'}{(2\pi)^4}\frac{M_W	^2}{M_W^2-q^{\prime 2}}\frac{\epsilon^{\mu\nu\alpha\lambda}q_\alpha'(T_{\mu\nu})_A}{[(p_\ell-q')^2-m_\ell^2]q^{\prime 2}}~.
\end{equation}
This integral is sensitive to the loop momentum $q'$ at all scales, from zero to infinity. For large values of $q'$ the integral can be computed precisely using pQCD, but at $q'\sim 1$~GeV the integrand becomes totally non-perturbative. Fortunately, recent improvements in lattice QCD techniques allow us to compute $T_{\mu\nu}$ from first principles in the forward limit ($p_i=p_f$) at small $q'$~\cite{Feng:2020zdc,Ma:2021azh}, so the combination of lattice QCD and pQCD yields a precise prediction of $I_\mathfrak{B}^\lambda$; the uncertainty from the non-forward (NF) corrections can be estimated with chiral power counting.

Finally, the contribution from $\delta F_3^\lambda$ involves a three-current product which is extremely difficult to compute on lattice, so we may resort to a fixed-order ChPT calculation. Nevertheless, we can improve the result by requiring the exact cancellation of the IR-divergence between the virtual correction and bremsstrahlung processes, and that allows us to resum the numerically-large IR-divergent piece to all orders. This amounts to writing:
\begin{equation}
	\delta f_{+,3}^{K\pi}=(\delta f_{+,3}^{K\pi})^{\text{IR}}+\left\{(\delta f_{+,3}^{K\pi})^{\text{fin}}_{e^2p^2}+\mathcal{O}(e^2p^4)\right\}~,
\end{equation}
where $(\delta f_{+,3}^{K\pi})^{\text{IR}}$ is the IR-divergent piece that is exactly known, while $(\delta f_{+,3}^{K\pi})^{\text{fin}}_{e^2p^2}$ is the IR-finite piece computed to $\mathcal{O}(e^2p^2)$ in ChPT. In fact, we found that the latter is reabsorbed into the tree-level form factor and isospin-breaking corrections according to the standard ChPT classification~\cite{Cirigliano:2011ny}, and is thus not included as a part of $\delta_\text{EM}^{K\ell}$. 

\subsection{$\delta f_-^{K\pi}$} 

Due to the $m_\ell^2/M_K^2$ suppression, one needs to compute $\delta f_-^{K\pi}$ only for the $K\rightarrow \pi\mu\nu$ channels, which was performed in Ref.\cite{Seng:2022wcw}. We express the result as
\begin{equation}
	\delta f_-^{K\pi}=(\delta f_-^{K\pi})_\text{IR}+(\delta f_-^{K\pi})_\text{conv}^{\text{fin}}+(\delta f_-^{K\pi})_\text{rem}~.
\end{equation}
The first two terms at the right hand side are analytically-known, IR-divergent and finite pieces coming from the first line in Eq.\eqref{eq:Sirlinmain} as well as the convection term contribution to the second line; they are the numerically largest pieces. On the other hand, $(\delta f_-^{K\pi})_\text{rem}$ denotes the remaining, numerically smaller piece which we compute at fixed order in ChPT, and is therefore subject to the usual chiral expansion and LEC uncertainties, but in a suppressed manner. 

\subsection{Bremsstrahlung}
 
\begin{figure}
	\centering
	\includegraphics[width=0.5\columnwidth]{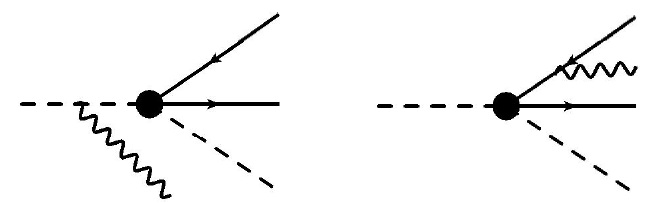}
	\caption{Bremsstrahlung diagrams.\label{fig:brem}} 
\end{figure} 

To cancel the IR divergence in the loop diagrams, one needs to include the real photon emission contribution at tree level, depicted in Fig.\ref{fig:brem}. The corresponding amplitude reads:
\begin{eqnarray}
	\mathcal{M}_\text{brem}&=&-\frac{G_Fe}{\sqrt{2}}\bar{u}_\nu\gamma^\mu(1-\gamma_5)\left\{\frac{p_\ell\cdot\varepsilon^*}{p_\ell\cdot q'}+\frac{\slashed{q}'\slashed{\varepsilon}^*}{2p_\ell\cdot q'}\right\}v_\ell F_\mu\nonumber\\
	&&+\frac{iG_Fe}{\sqrt{2}}\bar{u}_\nu\gamma^\nu(1-\gamma_5)v_\ell \varepsilon^{\mu *}T_{\mu\nu}~,\label{eq:Mbrem}
\end{eqnarray}
where $q'$ here denotes the on-shell outgoing photon momentum, and $\varepsilon^{\mu*}$ is the outgoing photon polarization vector. One sees that the first term at the right hand side is fixed by the experimental weak form factors, and all the theory uncertainties resides in the second term that contains $T_{\mu\nu}$. To separate the IR-divergent and IR-finite part of the amplitude, we can split:
\begin{equation}
	T^{\mu\nu}=T_\text{conv}^{\mu\nu}+\left\{(T^{\mu\nu}-T_\text{conv}^{\mu\nu})_{p^2}+\mathcal{O}(p^4)\right\}~,
\end{equation} 
and correspondingly, $\mathcal{M}_\text{brem}=\mathcal{M}_A+\mathcal{M}_B$, where $\mathcal{M}_A$ sums the first term and the $T^{\mu\nu}_\text{conv}$ contribution to the second term in Eq.\eqref{eq:Mbrem}; it contains all the IR-divergence but is exactly known. $\mathcal{M}_B$, on the other hand, represents the remaining piece and is expanded to the leading order in ChPT, and hence carries a ChPT expansion uncertainty. Requiring the IR-divergence in $|\mathcal{M}_A|^2$ cancels exactly with that in the loop contribution helps resumming the IR-divergent part of $\delta F_3^\lambda$, as we described in the previous section.

\subsection{Final result of $\delta_\text{EM}^{K\ell}$ and $V_{us}$}

\begin{table}
	\begin{centering}
		\begin{tabular}{cc}
			\hline 
			Mode & $\delta_\text{EM}^{K\ell}(\%)$\tabularnewline
			\hline 
			$K_{e3}^0$ & $11.6(2)_{\text{inel}}(1)_{\text{lat}}(1)_{\text{NF}}(2)_{e^{2}p^{4}}$\tabularnewline
			\hline 
			$K_{e3}^+$ & $2.1(2)_{\text{inel}}(1)_{\text{lat}}(4)_{\text{NF}}(1)_{e^{2}p^{4}}$ \tabularnewline
			\hline 
			$K_{\mu 3}^0$ & $15.4(2)_{\text{inel}}(1)_{\text{lat}}(1)_{\text{NF}}(2)_{\text{LEC}}(2)_{e^{2}p^{4}}$ \tabularnewline
			\hline 
			$K_{\mu 3}^+$ & $0.5(2)_{\text{inel}}(1)_{\text{lat}}(4)_{\text{NF}}(2)_{\text{LEC}}(2)_{e^{2}p^{4}}$\tabularnewline
			\hline 
		\end{tabular}
		\par\end{centering}
	\caption{\label{tab:deltaEM}Our new determination of $\delta_{\text{EM}}^{K\ell}$.}
	
\end{table}

\begin{table}
	\begin{centering}
		\begin{tabular}{|c|c|}
			\hline 
			& $|V_{us}f_{+}^{K^0\pi^-}(0)|$\tabularnewline
			\hline 
			\hline 
			$K_{e3}^L$ & $0.21617(46)_{\text{exp}}(10)_{I_{K}}(4)_{\delta_{\text{EM}}}(3)_{\text{HO}}$\tabularnewline
			\hline 
			$K_{e3}^S$ & $0.21530(122)_{\text{exp}}(10)_{I_{K}}(4)_{\delta_{\text{EM}}}(3)_{\text{HO}}$\tabularnewline
			\hline 
			$K_{e3}^+$ & $0.21714(88)_{\text{exp}}(10)_{I_{K}}(21)_{\delta_{\text{SU(2)}}}(5)_{\delta_{\text{EM}}}(3)_{\text{HO}}$\tabularnewline
			\hline 
			$K_{\mu 3}^L$ & $0.21649(50)_{\text{exp}}(16)_{I_{K}}(4)_{\delta_{\text{EM}}}(3)_{\text{HO}}$\tabularnewline
			\hline 
			$K_{\mu 3}^S$ & $0.21251(466)_{\text{exp}}(16)_{I_{K}}(4)_{\delta_{\text{EM}}}(3)_{\text{HO}}$\tabularnewline
			\hline 
			$K_{\mu 3}^+$ & $0.21699(108)_{\text{exp}}(16)_{I_{K}}(21)_{\delta_{\text{SU(2)}}}(6)_{\delta_{\text{EM}}}(3)_{\text{HO}}$\tabularnewline
			\hline 
			Average: $Ke$ & $0.21626(40)_{K}(3)_{\text{HO}}$\tabularnewline
			\hline 
			Average: $K\mu$ & $0.21654(48)_{K}(3)_{\text{HO}}$\tabularnewline
			\hline 
			Average: tot & $0.21634(38)_{K}(3)_{\text{HO}}$\tabularnewline
			\hline 
		\end{tabular}
		\par\end{centering}
	\caption{\label{tab:final}$|V_{us}f_{+}^{K^0\pi^-}(0)|$ from different $K_{\ell3}$ channels. ``HO'' denotes the uncertainty from the higher-order electroweak RC that resides in $S_\text{EW}$. }

\end{table}

Our final result of $\delta_\text{EM}^{K\ell}$ is summarized in Table~\ref{tab:deltaEM}. The sources of uncertainties are:
\begin{enumerate}
	\item inel: Inelastic states contribution to the residual integral
	\item lat: Lattice uncertainty in $I_\mathfrak{B}^\lambda$
	\item NF: Non-forward effects in $I_\mathfrak{B}^\lambda$
	\item $e^2p^4$: Higher-order ChPT corrections
	\item LEC: Poorly-determined LECs in $\delta f_-^{K\pi}$
\end{enumerate}
The results in Table~\ref{tab:deltaEMChPT} and \ref{tab:deltaEM} are in good agreement, but the latter has a significant improvement in precision from $10^{-3}$ to $10^{-4}$. The resultant product $|V_{us}f_{+}^{K^0\pi^-}(0)|$ are summarized in Table~\ref{tab:final}, where we observe a good consistency between different channels. Applying the lattice result of $f_+^{K^0\pi^-}(0)$ in Eq.\eqref{eq:latticef}, we obtain:
\begin{eqnarray}
	|V_{us}|_{K_{\ell3}} & = & \left\{ \begin{array}{ccc}
		0.22308(39)_{\text{lat}}(39)_{K}(3)_{\text{HO}} &  & N_{f}=2+1+1\\
		0.22356(62)_{\text{lat}}(39)_{K}(3)_{\text{HO}} &  & N_{f}=2+1
	\end{array}\right.~.
\end{eqnarray}
The result above can then be used, together with $V_{ud}$ obtained from superallowed nuclear decays and $V_{us}/V_{ud}$ obtained from kaon/pion leptonic decays, to test the validity of the Cabibbo unitarity. A global fit of several best determinations of $V_{ud}$ and $V_{us}$ indicates a $2.8~\sigma$ deficit from the Cabibbo unitarity~\cite{Cirigliano:2022yyo}, which may point towards new physics such as right-handed quark couplings~\cite{Cirigliano:2023nol}. This anomaly was first unveiled in the recent years following a re-analysis of the RC to the free neutron decay~\cite{Seng:2018yzq,Seng:2018qru}, and our high-precision calculation of the $K_{\ell 3}$ RC serves to further sharpen its status. In addition, there is a long-standing discrepancy between the determination of $V_{us}$ from kaon/pion leptonic decay + superallowed nuclear decays and from $K_{\ell 3}$ directly~\cite{ParticleDataGroup:2024cfk}, and the speculation was that it could be due to some hidden systematic effects in the $K_{\ell 3}$ RC, but our work has largely rejected this guess.  

\section{Summary}

To briefly summarize this talk:
\begin{itemize}
	\item The first-row CKM unitarity, often simplified as the Cabibbo unitarity, offers precision tests of the SM and sensitive probes for BSM physics competitive to high-energy collider experiments. The test requires precise extractions of $V_{ud}$ and $V_{us}$ from weak decays, and $K_{\ell 3}$ is the primary avenue for the latter.
	\item A number of SM theory inputs are needed for the extraction of $V_{us}$ from $K_{\ell 3}$ partial decay rate, including the process-independent electroweak RC, the $K^-\rightarrow\pi^0$ form factor, the phase space integral, the isospin-breaking correction, and the long-distance EM correction $\delta_\text{EM}^{K\ell}$.
	\item The previous state-of-the-art calculation of $\delta_\text{EM}^{K\ell}$ was performed using ChPT and reached a $10^{-3}$ precision; the main sources of uncertainties are the neglected higher-order terms in the chiral expansion and the poorly-known LECs.
	\item We performed a re-analysis of the $K_{\ell 3}$ EM correction by combining Sirlin's representation and ChPT approaches. The use of kaon and pion form factors and the lattice results of $K\pi$ box diagram overcome the limitations in the pure ChPT treatment and improve the precision level to $10^{-4}$.  
	\item Our work helps to consolidate the $V_{us}$ extraction, and sharpen the so-called ``Cabibbo angle anomaly'' since its first observation in 2018. 
\end{itemize} 

\section{Acknowledgments}

I thank Xu Feng, Daniel Galviz, Mikhail Gorchtein, Lu-Chang Jin, Peng-Xiang Ma, William J. Marciano and Ulf-G. Mei{\ss}ner for collaborations in related topics.
This work is supported in part by the U.S. Department of Energy (DOE), Office of Science, Office of Nuclear Physics, under award DE-FG02-97ER41014, and by the FRIB Theory Alliance award DE-SC0013617.

\bibliographystyle{JHEP}
\bibliography{ref}



\end{document}